\documentclass[prd,twocolumn,floats,superscriptaddress,eqsecnum,floatfix]{
revtex4}

\usepackage{amssymb,amsmath}
\usepackage{epsfig}

\newcommand{\be}{\begin{equation}}
\newcommand{\ee}{\end{equation}}
\newcommand{\ba}{\begin{eqnarray}}
\newcommand{\ea}{\end{eqnarray}}

\newcommand{\eq}[1]{Eq.(\ref{#1})}
\newcommand{\n}[1]{\label{#1}}

\newcommand{\BM}[1]{{\mbox{\boldmath $#1$}}}
\newcommand{\tp}{\tilde{\varphi}}

\newcommand{\hh}{\, ,\hspace{0.8cm}}
\newcommand{\hhh}{\, ,\hspace{0.4cm}}

\newcommand{\ins}[1]{{\mbox{\tiny #1}}}
\newcommand{\ind}[1]{{\mbox{\scriptsize #1}}}

\begin{document}

\title{Self-energy of a scalar charge near higher-dimensional  black holes}

\author{Valeri P. Frolov}%
\email[]{vfrolov@ualberta.ca}
\affiliation{Theoretical Physics Institute, Department of
Physics,
University of Alberta,\\
Edmonton, Alberta, Canada T6G 2E1
}
\author{Andrei Zelnikov}%
\email[]{zelnikov@ualberta.ca}
\affiliation{Theoretical Physics Institute, Department of Physics,
University of Alberta,\\
Edmonton, Alberta, Canada T6G 2E1
}


\begin{abstract}

We study the problem of self-energy of charges in higher dimensional
static spacetimes. Application of regularization methods of quantum field theory
to calculation of the classical self-energy of charges  leads to
model-independent results.
The correction to the self-energy of a scalar charge due to the gravitational
field of black holes of the higher dimensional Majumdar-Papapetrou
spacetime is calculated exactly. It proves to be zero in even dimensions,
but it acquires non-zero value in odd dimensional spacetimes. The origin of the
self-energy correction in odd dimensions is similar to the origin the conformal
anomalies in quantum field theory in even dimensional spacetimes.
\end{abstract}


\maketitle

\section{Introduction}

There are several problems in the theoretical physics that have quite long story
but still attract a lot of attention.
The problem of the electromagnetic origin of the electron mass is one of them.
It first was formulated in the classical theory when in 1881 Thompson
\cite{Thomson:1881} demonstrated that the self-energy of the electromagnetic
field contributes to the inertial mass of a charged particle. This idea was then
elaborated in the works by Lorentz \cite{Lorentz:1899,Lorentz:1904},
Abraham \cite{Abraham:1903}, Poincar\'{e} \cite{Poincare:1905},
Fermi \cite{Fermi:1921} and other. For a simple model of a uniformly charged
sphere
of radius $\varepsilon$, the electrostatic energy is $E=e^2/(2\varepsilon)$.
However it was shown by Abraham \cite{Abraham:1904,Abraham:1905} the relation
between energy and momentum for such a particle differs from the standard one by
a factor 4/3. This factor disappears if one includes in the definition of the
self-energy a contribution of additional (non-electromagnetic) forces that are
required to make a system stable. To solve 4/3-problem Poincar\'{e}
\cite{Poincare:1905,Poincare:1906} introduced special sort of
non-electromagnetic pressure. Max von Laue \cite{Laue:1911} formulated a general
theorem, demonstrating that whenever a spatially extended system is stable, the
total mass of such a system $m_\ins{tot}$ is always related to its rest-mass as
follows $m_\ins{tot}=E_\ins{tot}/c^2$.
The problem of self-energy and stability of a classical electron is discussed in
many more recent papers  (see e.g.
\cite{Rohrlich:1960,Tangherlini:1962,Boyer:1982,Blinder:2001,Blinder:2008}).

There are many different ways how a simple spherical-shell model of a classical
electron can be modified.
For example, instead of the shell one can consider a charged ball, the shape of
the shell or of the ball can be deformed, the distribution of the electric
charge can be non homogeneous, and so on. Certainly, to be consistent, for each
of these modifications one must also modify the non-electromagnetic forces in
order to satisfy the Laue's theorem. Bopp \cite{Bopp:1940} and Podolsky
\cite{Podolsky:1942,PodolskySchwed:1948} proposed a covariant scheme for the
calculation of the
classical self-energy of the electron. Their idea was to start with a
higher-derivative modification of the corresponding field theory. For example,
for a scalar massless field $\varphi$ one starts with the equation
\be\nonumber
(1-{1\over \mu^2}\Box) \Box \varphi=-4\pi J\, .
\ee
This equation is equivalent to a set of relations
\be\nonumber
 \varphi= \varphi'- \varphi''\,,
\ee
\be\nonumber
 \varphi'= (1-{1\over \mu^2}\Box) \varphi\hh
\varphi''= -{1\over \mu^2}\Box \varphi\, ,
\ee
if the fields $\varphi'$ and  $\varphi''$ obey the equations
\be\nonumber
\Box \varphi'=-4\pi J\hh  (\Box-\mu^2) \varphi''=-4\pi J\, .
\ee
For a point-like charge $q$ the infinite parts in the self-energy for both fields $\varphi'$ and
$\varphi''$ are identical,
and as result of their subtraction the self-energy  is
finite $E_\ins{self}=q^2 \mu$. In this regularization $\mu$ plays the role of the cut-off parameter.
In fact, in such an approach
for a small size of a classical charged particle $\varepsilon\ll \mu^{-1}$  all
the details of the charge distribution become unimportant.
One can easily see that this approach has many  common features with the
Pauli-Villars regularization widely used in the modern quantum field
theory\footnote{In the standard theory the bare mass of elementary particles is
explained by the Higgs mechanism. However the self-interaction gives
contribution to the renormalization of the mass.}

In the quantum electrodynamics the self-energy of the electron is
divergent in the point-particle limit. In the second order of the perturbation
theory this divergence is of the form
\be\n{qse}
\Delta m \sim m_0 {3 e^2\over 2\pi \hbar c}\ln (\hbar/ (\varepsilon m_0 c))\, ,
\ee
where $m_0$ is the `bare' mass of the electron, and $\varepsilon$ is the cut-off
radius.
In the limit $\hbar\to 0$ the expression \eq{qse} does not reproduce the
classical result. The reason of this is that in order to derive this relation one
uses the expansion in $\alpha=e^2/(\hbar c)$.
However, as it was demonstrated by Vilenkin and Fomin
\cite{Vilenkin_Fomin:1974,Vilenkin:1979},  there exists a correct
quantum-to-classical correspondence for the self-energy of the electron
(see also \cite{EfimovIvanovMogilevsky:1977}.)

In the presence of the gravitational field the self-energy problem becomes more
complicated. The reason is that the field of a pointlike charge is spread in
space, and its contribution to the energy in a general case is non-local.
The classical
fields created by charges are not localized at
the position of the charge. It means that the charges are to be treated
as extended objects. Qualitatively the origin of the classical electromagnetic
self-force of charges can
be explained by the deformation of the distribution of their classical fields
in curved spacetime that leads to an extra force acting on the charge
itself.
Fermi \cite{Fermi:1921} showed that for a special case of the homogeneous static
gravitational field the self-energy is the same as in the absence of the field.
This result
can be related to the equivalence principle. However, in a general case
the electromagnetic (or scalar, or any other field)
self-energy depends on the position of the particle. This
may lead to an extra force acting on a charged particle.

In a generic case the self-force acting on a particle {\em moving} in the
gravitational field contains both conservative and dissipative
terms, the dissipative terms being responsible for the radiation reaction.
The fundamental problem of calculation of the radiation reaction of particles in
the external gravitational field \cite{DeWittBrehme:1960} has got much attention
\cite{MinoSasakiTanaka:1997,QuinnWald:1997,Quinn:2000} in
connection with the study of waveforms of gravitational radiation, especially
because of possible applications in experiments for gravitational wave
detection. A nice review on the contemporary state of the problem one can find
in \cite{Poisson:2004}.

For a static particle in a static (stationary) gravitational field the radiation
force
is absent, the problem is simplified, and in some special cases the self-energy
can be calculated exactly.
This becomes possible when the static Green functions of classical fields are
known
exactly. Fortunately this is the case for some physically interesting systems
like static charge near
4-dimensional Schwarzschild or Reissner-Nordstr\"{o}m black holes
\cite{Copson:1928,LeauteLinet:1976,Linet:1976}.
For the electric pointlike charge an
self-interaction energy is
\cite{SmithWill:1980,FrolovZelnikov:1982,Lohiya:1982}
\be\begin{split}
E&=\left(m_\ins{\,bare}+{e^2\over 2
\epsilon}\right) \,|g_{00}|^{1/2}+\Delta E\,,\\
\Delta E&={e^2 M\over 2 r^2}\,,
\end{split}\ee
which leads to an additional repulsive (directed from the black hole)
self-force.
Here $\epsilon$ is the classical radius
of
the electron, $m_\ind{bare}$ is its bare mass, and $r$ is the 
radial distance to the black hole.
It was also demonstrated that for a  scalar charge
near the Reissner-Nordstr\"{o}m black hole the self-energy correction $\Delta E$
 vanishes
\cite{Quinn:2000,ChoTsokarosWisseman:2007}.
In \cite{BezerraKhusnutdinov:2009} the self-energy of scalar charges in the
background geometry of wormholes was studied.

The aim of this paper is to analyze the self-energy problem for a static
pointlike charge in a static higher-dimensional spacetime.
Our motivations for this analysis are the following. In higher dimensions the
fields near a pointlike source is stronger than in 4D case. Hence one can expect
much more dramatic  dependence of the divergent part of the classical
self-energy on details of a model of a classical source. We demonstrate that
for the calculation of the self-energy one can use methods similar to the ones
adopted in the quantum field theory. To be concrete we focus on the
point-splitting method. It is well known for the calculations of the quantum
vacuum polarization effects in a curved spacetime. We demonstrate that this
regularization method  works
well for the calculation of the self-energy.

Another interesting question is why non-local self-energy
correction $\Delta E$ for a scalar massless field vanishes in 4D. We shall
demonstrate that this is a generic property of a wide class of even-dimensional
spacetimes with a
spatial metric conformal to the flat one, while in odd
dimensional case there exists a non-vanishing extra force acting on a charged
scalar source.

The paper is organized as follows.
In Section~II we discuss the self-energy problem for a scalar charge in a
static 
gravitational field and obtain expression for the corresponding shift of mass in
terms of a static Green function.
After this we adapt the point-splitting formalism, well known in the quantum
field
theory, for the classical self-energy problem in a general higher dimensional
static gravitational field. This approach allows one (at least formally) to
avoid problems connected with the details of the charge structure (Section
\ref{Point-splitting}).
We illustrate this method by calculations of the self-energy of a scalar
massless charge.
We apply the point-splitting approach for the calculation of the self-energy of
a static source in a 4D static black hole metrics and
show that this formalism correctly reproduces known 4D results for the
self-force
(Section \ref{Schwarzschild}). In Section~V we use the point-splitting method for
the calculation
of the self-energy of pointlike scalar charges at rest in the vicinity of
higher-dimensional extremely charged black hole, or a set of such black holes.
For such gravitational backgrounds the exact static Green functions are known
\cite{FrolovZelnikov:2012}, so that one is able to obtain an exact explicit
expression for the
self-energy. We demonstrate that in even dimensional Majumdar-Papapetrou
spacetimes the self-force vanishes, while in the odd dimensional ones the
self-force and self-energy can be related with conformal anomalies. Conclusions
contain a discussion of the obtained results.


\section{Self-energy of a scalar charge in a static
spacetime}\label{Self-energy}

Let us consider a scalar massless field $\varphi$ in $D-$dimensional spacetime
with metric
\be
dS^2=g_{\mu\nu}dy^{\mu}dy^{\nu}\, .
\ee
It obeys the equation
\be
\Box\varphi\equiv -4\pi J\, .
\ee
We assume that the spacetime is static and $\BM{\xi}$ is its Killing vector, so
that in the region where $\alpha^2\equiv -\BM{\xi}^2>0$ one can write the metric
in the form
\be
dS^2=-\alpha^2 dt^2+g_{ab}dx^a dx^b       \hhh
\partial_t\alpha=\partial_t g_{ab}=0\, .
\ee
For a static source the field equation takes the form
\be\n{eq1}
\triangle \varphi +(\nabla \ln\alpha,\nabla\varphi)=-4\pi J\, .
\ee
Here $\nabla_a$ is a covariant derivative in $(D-1)-$dimensional metric $g_{ab}$
and
$\triangle=g^{ab}\nabla_a\nabla_b$.

For a pointlike scalar charge $q$ located at $\BM{x}$ one has
\be\begin{split}
J(\BM{x})&=q\int_{-\infty}^{\infty} d\tau \delta^{D-1}(\BM{x},\BM{x}')
{\delta(t-t'(\tau))\over \alpha(\BM{x}')}\\
&=q \,\delta^{D-1}(\BM{x},\BM{x}')\, ,\\
\delta^{D-1}(\BM{x},\BM{x}') &={ \delta^{D-1}(\BM{x}-\BM{x}')\over \sqrt{g}}\hhh
g=\det g_{ab} \,.
\end{split}\ee

In what follows it is convenient to rewrite \eq{eq1} in a  self-adjoint form.
For this purpose we introduce the quantities
\be
\varphi=\alpha^{-1/2}\tilde{\varphi}\hh J=\alpha^{-1/2}j\, ,
\ee
and write the \eq{eq1} in the form
\be\begin{split}\n{eq2}
\hat{F}\tp&\equiv (\triangle  +V)\tp=-4\pi j\,,\\
V&={(\nabla\alpha)^2\over
4\alpha^2}-{\triangle\alpha\over 2\alpha}\equiv -{\triangle(\alpha^{1/2})\over
\alpha^{1/2}} \, .
\end{split}\ee

The energy $E$ in a static spacetime is
\be
E=\int_{\Sigma} T_{\mu\nu}\xi^{\mu}d\Sigma^{\nu},
\ee
where $\Sigma$ is a Cauchy surface and $d\Sigma^{\nu}$ is a future-directed
volume element on it. The energy-momentum tensor for the
minimally coupled massless scalar field is
\be
T_{\mu\nu}={1\over 4\pi}\left( \varphi_{;\mu}\varphi_{;\nu}-{1\over
2}g_{\mu\nu}\varphi_{;\alpha}\varphi^{;\alpha}
\right)+g_{\mu\nu}\varphi J \,.
\ee
For a static field $\xi^{\mu}\varphi_{,\mu}=0$ so that one has
\be
E=\int_{\Sigma} {\cal T}\xi_{\mu}d\Sigma^{\mu}\hh
{\cal T}=J\varphi-{1\over 8\pi}(\nabla\varphi)^2\, .
\ee
Since $E$ does not depend on the choice of $\Sigma$, we chose this surface in
the form $t=$const. In the presence of a black hole one restricts the integration domain by the black hole exterior. For this choice of $\Sigma$ one has
\be
d\Sigma^{\mu}=n^{\mu}\sqrt{g}\,d^{D-1}x\, ,
\ee
where $n^{\mu}$ is a unit future-directed vector normal to $\Sigma$, and
\be
\xi_{\mu}n^{\mu}=-\alpha\, .
\ee
Thus
\be
E=-\int_{t=\ind{const}}d^{D-1}x \sqrt{g}\,\alpha {\cal T}\, .
\ee
Using the Stock's theorem and the field equation \eq{eq1} we get
\footnote{Notice that the boundary term at infinity in the Stock's formula
vanished since the field $\tp$ decreases there rapidly enough. In the presence
of a black hole there exists also a boundary term at its horizon. However for a
static regular field it vanishes as well (for details see e.g.
\cite{FrolovZelnikov:1982})}
\be\n{EE}
E=-{1\over 2}\int_{\Sigma}\alpha\varphi J\sqrt{g}\,d^{D-1}x
= -{1\over 2}\int_{\Sigma}\tp j\sqrt{g}\,d^{D-1}x \,.
\ee

Denote by ${\cal G}$ the Green function of the
operator ${\hat F}$
\be\n{FG}
\hat{F}\,{\cal G}(\boldsymbol{x},\boldsymbol{x}')=-4\pi
\delta^{D-1}(\BM{x},\BM{x}')\, .
\ee
Then \eq{EE} takes the form
\be
E=-{q^2\over 2}\alpha(\BM{x}){\cal G}(\boldsymbol{x},\boldsymbol{x})\, .
\ee
As expected, the obtained expression for the self-energy of
a pointlike charge is divergent.
To deal with this problem we shall use the point-splitting method, similar to
the regularization schemes  adopted in the quantum field
theory. Namely, to regularize $E$ we use the regularized
version of the Green function ${\cal
G}_\ins{reg}(\boldsymbol{x},\boldsymbol{x})$
\be
{\cal
G}(\boldsymbol{x},\boldsymbol{x})\rightarrow
{\cal G}_\ins{reg}(\boldsymbol{x},\boldsymbol{x})=\lim_{\boldsymbol{x}
\rightarrow\boldsymbol{x}'}
\left[{\cal G}(\boldsymbol{x},\boldsymbol{x}')-{\cal
G}_\ins{div}(\boldsymbol{x},\boldsymbol{x}')\right]\, .
\ee
We discuss this point-splitting procedure in the next section. Now let us make
the following remark. The energy of the object of mass $m$ at rest at a point
$\BM{x}$ in a static gravitational is
\be
E=-m u_{\mu}\xi^{\mu}=m\, \alpha(\BM{x})\, .
\ee
Using this relation we obtain for the contribution $\Delta m$ of the self-energy
to the mass of a scalar charge the following expression
\be
\Delta m=-{q^2\over 2} {\cal G}_\ins{reg}(\boldsymbol{x},\boldsymbol{x})\, .
\ee

\section{Point-splitting regularization of of the
self-mass}\label{Point-splitting}

\subsection{Schwinger--DeWitt expansion}

To obtain ${\cal G}_\ins{div}$ it is convenient to start with the heat kernel
expansion for the operator $\hat{F}$. We define the heat kernel
${\cal K}(s|\boldsymbol{x},\boldsymbol{x}')$ as a solution of the equation
\be\begin{split}
\left[-{\partial\over \partial
s}+\hat{F}\right]\,{\cal
K}(s|\boldsymbol{x},\boldsymbol{x}')
&=-\delta^{D-1}(\boldsymbol{x}
,\boldsymbol{x}')\,\delta(s)\, .
\end{split}\ee
The static Green function ${\cal G}(\boldsymbol{x},\boldsymbol{x}')$ defined by
\eq{FG} is
\be\label{calG}
{\cal G}(\boldsymbol{x},\boldsymbol{x}')
=\int_0^\infty ds\,{\cal K}(s|\boldsymbol{x},\boldsymbol{x}') \,.
\ee
The divergent terms of ${\cal G}$ are determined by
the behavior of the heat kernels at small $s$
and can be found by using the standard Schwinger--DeWitt expansion
\be\begin{split}\label{calK}
{\cal
K}(s|\boldsymbol{x},\boldsymbol{x}')&={\Delta^{1/2}
(\boldsymbol{x},\boldsymbol{x}')\over (4\pi
s)^{(n+2)/2}}\,\exp\left({-{{\sigma}(\boldsymbol{x},
\boldsymbol{x}
')\over 2 s}}\right)\,\\
&\times\sum_{k=0}^{\infty}
{a}_k(\boldsymbol{x},\boldsymbol{x}')s^k\,.
\end{split}\ee
Here ${a}_k$  are the Schwinger--DeWitt coefficients
for the operator $\hat{F}$. The world function
${\sigma}$ and Van Vleck--Morette determinant
$\Delta$ are defined on the $(D-1)-$dimensional spatial metric $g_{ab}$.

The divergent part of the static Green function
\be\label{calGdiv}
{\cal
G}_\ins{div}(\boldsymbol{x},\boldsymbol{x}')
=\int_0^\infty ds\,{\cal
K}_\ins{div}(s|\boldsymbol{x},\boldsymbol{x}')
\ee
comes from the first $[(D-1)/2]$ terms in this series \eq{calK}.
Denote $n=D-3$ then one has
\be\begin{split}\label{calKdiv}
{\cal
K}_\ins{div}(s|\boldsymbol{x},\boldsymbol{x}')&={\Delta^{
1/2}
(\boldsymbol{x},\boldsymbol{x}')\over (4\pi
s)^{(n+2)/2}}\,\exp\left({-{{\sigma}(\boldsymbol{x},
\boldsymbol{x}
')\over 2 s}}\right)\,\\
&\times\sum_{k=0}^{[n/2]}
{a}_k(\boldsymbol{x},\boldsymbol{x}')s^k\,.
\end{split}\ee
Therefore
\be\begin{split}\label{calGdiva}
{\cal
G}_\ins{div}(\boldsymbol{x},\boldsymbol{x}')
&={\Delta^{1/2}
(\boldsymbol{x},\boldsymbol{x}')
{1\over (2\pi )^{{n\over 2}+1}}}\,\\
&\times\sum_{k=0}^{[n/2]}
{\Gamma\left({{n\over 2}-k}\right)\over
2^{k+1}\sigma^{{n\over
2}-k}}
{a}_k(\boldsymbol{x},\boldsymbol{x}') .
\end{split}\ee
When $n$ is even the last term $(k=n/2)$ in the sum
should be substituted by
\be\begin{split}
{\Gamma\left({{n\over 2}-k}\right)\over
2^{k+1}{\sigma}^{{n\over
2}-k}}\,&
{a}_k(\boldsymbol{x},\boldsymbol{x}')\Big|_{k=n/2}\\
&\rightarrow
-{\ln{\sigma}(\boldsymbol{x},\boldsymbol{x}')+\gamma-\ln
2\over 2^{{n\over
2}+1}}\,{a}_{n/2}(\boldsymbol{x},\boldsymbol{x}') .
\end{split}\ee

\subsection{Special cases \label{sc}}

Let us illustrate the self-mass calculations by examples of
special static black hole  solutions in a spacetime with dimensions $D=4,5,6$,
and $7$. For this purpose let us present here the corresponding expressions for
the divergent parts of the static Green functions for these cases.

\begin{itemize}
                   \item{\bf  Four dimensions $D=4,n=1$}
\be\label{Gdiv4}
{\cal
G}_\ins{div}(\boldsymbol{x},\boldsymbol{x}')
=
{\Delta^{1/2}(\boldsymbol{x},\boldsymbol{x}')\over 4\pi}\,
{1\over
(2{\sigma})^{1/2}}{a}_0(\boldsymbol{x},\boldsymbol{x}
')\,.
\ee
                    \item{\bf  Five dimensions $D=5,n=2$}
\be\begin{split}\label{Gdiv5}
{\cal
G}_\ins{div}(\boldsymbol{x},\boldsymbol{x}')
&=
{\Delta^{1/2}(\boldsymbol{x},\boldsymbol{x}')\over
4\pi^2}\,\left[
{1\over
2{\sigma}}\,{a}_0(\boldsymbol{x},\boldsymbol{x}')\right.\\
 & \left. - {
1\over
4}(\ln{\sigma}+\gamma-\ln 2)\,{a}_1(\boldsymbol { x },
\boldsymbol{x}')\right] \,.
\end{split}\ee
                     \item{\bf  Six dimensions $D=6,n=3$}
\be\begin{split}\label{Gdiv6}
{\cal
G}_\ins{div}(\boldsymbol{x},\boldsymbol{x}')
=
{\Delta^{1/2}(\boldsymbol{x},\boldsymbol{x}')\over
8\pi^2}&\left[
{1\over
(2{\sigma})^{3/2}}\,{a}_0(\boldsymbol{x},\boldsymbol{x}') \right.\\ &\left.
+{1\over2(2{\sigma})^{1/2}}\,
{a}_1(\boldsymbol{x } , \boldsymbol{x}')\right] \,.
\end{split}\ee
                   \item{\bf  Seven dimensions $D=7,n=4$}
\be\begin{split}\label{Gdiv7}
{\cal
G}_\ins{div}(\boldsymbol{x},\boldsymbol{x}')&=
{\Delta^{1/2}(\boldsymbol{x},\boldsymbol{x}')\over
8\pi^3}\,\\
&\times \left[
{1\over
2{\sigma}^2}\,{a}_0(\boldsymbol{x},\boldsymbol{x}')
+{1\over 4{\sigma}}\,{ \alpha}_1(\boldsymbol{x } ,
\boldsymbol{x}') \right.\\
 & \left.
\hskip 0.5cm   -{1\over 8}(\ln{\sigma}+\gamma-\ln
2)\,{a}_2(\boldsymbol { x } ,
\boldsymbol{x}')
\right] \,.
\end{split}\ee
\end{itemize}

\section{Self-energy in four dimensional
Reissner-Nordstr\"om spacetime}\label{Schwarzschild}

As a first example let us apply the developed point-splitting method to
calculation of the self-energy of a
scalar charge  near a four-dimensional Reissner-Nordstr\"om
black hole. Namely, we shall demonstrate that this methods give he same answer
as earlier calculations using a spherical shell model of a the classical charged
particle.

Let $M$ and $Q$ be mass and electric charge
$Q$ of the lack hole. The Reissner-Nordstr\"om metric in the isotropic
coordinates is
\be
dS^2=-\alpha^2\,dt^2 + U^2 \delta_{ab}\,dx^adx^b\,,
\ee
where
\be\begin{split}
\alpha &={4\rho^2-(M^2-Q^2)\over(2\rho+M+Q)(2\rho+M-Q)} \,,\\
U&=1+{M\over \rho}+{M^2-Q^2\over 4\rho^2} \hh \rho^2=\delta_{ab}\,x^a
x^b\,.
\end{split}\ee
The standard radial coordinate $r$ is related to the
isotropic coordinate $\rho$  as follows
\be
r=M+\rho+{M^2-Q^2\over 4\rho}\,.
\ee

The spatial metric is conformally flat
\be\label{gab}
g_{ab}=U^2\delta_{ab}\,.
\ee
We denote a `coordinate distance' between two points as
\be
|\boldsymbol{x} -\boldsymbol { x }'|\equiv
\sqrt{\delta_{ab}(x^a-{x'}^a)(x^b-{x'}^b)}\,.
\ee
In spherical isotropic coordinates $(t,\rho,\theta,\phi)$
\be
dS^2=-\alpha^2\,dt^2+U^2\left[d\rho^2+\rho^2\left(d\theta^2+\sin^2\theta\,
d\phi^2\right) \right]
\ee
it takes the form
\be\begin{split}
&|\boldsymbol{x} -\boldsymbol { x
}'|=\sqrt{\rho^2+{\rho'}^2-2\rho\rho'\lambda}\,,\\
&\lambda=\cos{\theta}\cos{\theta'}+\sin{\theta}\sin{\theta'}
\cos{(\phi-\phi')}\,.
\end{split}\ee
The static Green function for a scalar field in the Reissner-Nordstr\"om
spacetime is known exactly \cite{Linet:1977,FrolovZelnikov:1980,Linet:2005}.
In  isotropic coordinates it takes the form
\be\begin{split}\label{G4exact}
{\cal G}(\boldsymbol{x},\boldsymbol{x}')
&={\sqrt{\alpha\alpha'}\over 4\pi R(\boldsymbol{x},\boldsymbol{x}')}
, \hskip 3mm
\alpha=\alpha(\rho) , \hskip 3mm
\alpha'=\alpha(\rho')\, .
\end{split}\ee
Here
\be\begin{split}
 R(\boldsymbol{x},\boldsymbol{x}')^2&=(\rho^2+{\rho'}^2-2\rho\rho'\lambda)\\
&\times\left [
1-{\lambda(M^2-Q^2)\over 2\rho\rho'}+{(M^2-Q^2)^2\over 16\rho^2{\rho'}^2}
\right]\,.
\end{split}\ee
In the limit $\BM{x}'\to \BM{x}$ one has the following expressions
\be\begin{split}
{\sigma}(\boldsymbol{x},\boldsymbol{x}')&={UU'\over
2}|\boldsymbol{x}-\boldsymbol{x}'|^2+O(|\boldsymbol{x}
-\boldsymbol { x }'|^4)\, ,\\
a_0(\boldsymbol{x},\boldsymbol{x}')&=1\, ,\\
\Delta^{1/2}(\boldsymbol{x},\boldsymbol{x}')&=1+O(|\boldsymbol{x}
-\boldsymbol { x }'|^2)\, .
\end{split}\ee
Thus \eq{Gdiv4} leads to
\be\label{Gdiv4a}
{\cal G}_\ins{div}(\boldsymbol{x},\boldsymbol{x}')
={1\over 4\pi\sqrt{UU'}}\,{1\over
|\boldsymbol{x} -\boldsymbol { x }'|}+O(|\boldsymbol{x}
-\boldsymbol { x }'|)\,.
\ee
Similarly, when $\boldsymbol{x}\rightarrow \boldsymbol{x}'$, one can expand
the exact expression \eq{G4exact} in series. It's easy to check that
\be
{\alpha\alpha'\over \left[
1-{\lambda(M^2-Q^2)\over 2\rho\rho'}+{(M^2-Q^2)^2\over
16\rho^2{\rho'}^2}\right]}={1\over UU' }+O(|\boldsymbol{x}
-\boldsymbol { x }'|^2)\, .
\ee
and, hence,
\be\label{G4a}
{\cal G}(\boldsymbol{x},\boldsymbol{x}')
={1\over 4\pi\sqrt{UU'}}\,{1\over
|\boldsymbol{x} -\boldsymbol { x }'|}+O(|\boldsymbol{x}
-\boldsymbol { x }'|)\,.
\ee
By comparing \eq{G4a} with \eq{Gdiv4a} one obtains
\be
{\cal G}_\ins{reg}(\boldsymbol{x},\boldsymbol{x}')={\cal
G}(\boldsymbol{x},\boldsymbol{x}')-{\cal
G}_\ins{div}(\boldsymbol{x},\boldsymbol{x}')=O(|\boldsymbol{x}
-\boldsymbol { x }'|)
\ee
and hence
\be
{\cal G}_\ins{reg}(x,x)=0\,.
\ee
The corresponding self-energy $E_\ins{self}$ and the mass correction $\Delta m$
vanish
\be
\Delta m=\alpha^{-1} E_\ins{self}=0\,.
\ee
This result coincides with the corresponding result for a spherical shell model
obtained earlier
\cite{FrolovZelnikov:1982,QuinnWald:1997,Quinn:2000,ChoTsokarosWisseman:2007}.
Let us emphasize that the point-splitting method not only easier
and simplify calculations, but, what is more important, it allows one to extract
the finite part of the self-energy without discussing details of the classical
charged particle model. It also can be used in arbitrary number of spacetime
dimensions. In order to illustrate the latter point we shall perform
calculations of the self-energy in special higher-dimensional black hole
metrics.


\section{Self-energy of a scalar charge in the higher dimensional
Majumdar-Papapetrou
metrics}\label{Majumdar-Papapetrou}

\subsection{Static Green function in the Majumdar-Papapetrou spacetime}

There exist a wide class of higher dimensional metrics where the static Green
functions for scalar and electromagnetic field of a point charge are known in
explicit form \cite{FrolovZelnikov:2012}. These are so called higher dimensional
Majumdar-Papapetrou metrics \cite{Myers:1986}. They describe the field of a set
of extremely
charged black holes in equilibrium in a higher dimensional asymptotically flat
spacetime.
The corresponding background metric and electric potential are of the form
$(D=n+3$)
\be\begin{split}\label{MP}
dS^2&=-U^{-2}\,dt^2+U^{2/n}\delta_{ab}\,dx^a dx^b\, ,\\
A_{\mu}&=\sqrt{{n+1\over 2n}}\,U^{-1}\,\delta^0_{\mu}\,.
\end{split}\ee
Denote
\ba
\rho&=&|\BM{x}-\BM{x'}|=\sqrt{\delta_{ab}(x^a-{x'}^a)(x^b-{x'}^b)}\, ,\\
\BM{\triangle}&=&\delta^{ab}\partial_a\partial_b\, .
\ea
Note that the flat Laplace operator $\BM{\triangle}$ differs from the curved
one ${\triangle}=g^{ab}\nabla_a\nabla_b$.

The function $U$ in \eq{MP}
\be
U=1+\sum_k {M_k\over \rho_k^n}\hh
\rho_k=|\boldsymbol{x}-\boldsymbol{x}_k|\, .
\ee
The index $k=(1,\dots,N)$ enumerates the extremal black holes.
$x^a_k$ is the spatial position of the $k$-th extremal black
hole. The potential $U$ obeys the equation
\be\label{lapU}
\BM{\triangle} U =-{4\pi^{1+{n\over 2}}\over \Gamma\left({n\over
2}\right)}\sum_k
M_k\,\delta^{n+2}(\boldsymbol{x}-\boldsymbol{x}_k)\, .
\ee
That is  in the black holes exterior the function $U$ is a harmonic function

It is easy to heck that the static scalar field equation \eq{eq1} in the metric
\eq{gab} takes the form
\be
\BM{\triangle}\varphi=-4\pi U^{2/n}J\, .
\ee

For a pointlike charge this equation can be easily solved. It is sufficient to
use the following relations
\be\begin{split}
\BM{\triangle} \left[{1\over \rho^n}\right]&=-{4\pi^{1+{n\over 2}}\over
\Gamma\left({n\over
2}\right)}\,\delta^{n+2}(\boldsymbol{x}-\boldsymbol{x}')\\
&=-n{1\over \rho^{n+1}}\,\delta(\rho)\, ,
\end{split}\ee
\be\begin{split}
\delta^{n+2}(\boldsymbol{x}-\boldsymbol{x}')&={\Gamma\left(1+{n\over
2}\right)\over
2\pi^{1+{n\over 2}}}\,{1\over \rho^{n+1}}\,\delta(\rho)\, .
\end{split}\ee
The static Green function is \cite{FrolovZelnikov:2012}
\be\label{Gscalar}
{\cal G}(\boldsymbol{x},{\boldsymbol{x}'})={\Gamma\left({n\over
2}\right)\over
4\pi^{1+{n\over 2}}}
\cdot{1\over\sqrt{UU'}}\,{1\over \rho^n}\, .
\ee

\subsection{Self-energy}

Since the spatial part of the Majumdar-Papapetrou metric is conformally flat,
the calculations of the divergent part of a static function in higher
dimensional case are greatly simplified. In addition in this case the operator
$\hat{F}$ happens to be conformally invariant. The details of the calculations
can
be found in the Appendices. Using these results and expressions for ${\cal
G}_\ins{div}$ presented in the subsection \ref{sc}, one obtains $E_\ins{self}$
and
$\Delta m=E_\ins{self}/\alpha$.
Here we collect the corresponding results for $D=4,$ 5, and 6  dimensional
spacetimes

\begin{itemize}
                   \item{\bf Four dimensions $D=4,n=1$}
\be\begin{split}
{\cal
G}(\boldsymbol{x},\boldsymbol{x}')&={1\over 4\pi}{1\over
\sqrt{UU'}}\,{1\over|\boldsymbol{x}-\boldsymbol{x}'|}\, ,\\
{\cal
G}_\ins{div}(\boldsymbol{x},\boldsymbol{x}')&={1\over 4\pi}{1\over
\sqrt{UU'}}\,{1\over|\boldsymbol{x}-\boldsymbol{x}'|}+O(|\boldsymbol{x}
-\boldsymbol{x}'|)\, ,\\
{\cal
G}_\ins{reg}(\boldsymbol{x},\boldsymbol{x})&=0\, ,
\end{split}\ee
\be
\Delta m=0\,.
\ee
                    \item{\bf  Five dimensions $D=5,n=2$}
\be\begin{split}
{\cal G}(\boldsymbol{x},\boldsymbol{x}')
&={1\over 4\pi^2}{1\over
\sqrt{UU'}}\,{1\over|\boldsymbol{x}-\boldsymbol{x}'|^2}\, ,\\
{\cal G}_\ins{div}(\boldsymbol{x},\boldsymbol{x}')
&={1\over 4\pi^2}{1\over  \sqrt{UU'}}\,
\Big[{1\over
|\boldsymbol{x}-\boldsymbol{x}'|^2}  \\
&\hskip 1.7cm +{1\over 72}U{\cal R}
\Big]
+O(|\boldsymbol{x}-\boldsymbol{x}'|^2)\, ,\\
{\cal G}_\ins{reg}(\boldsymbol{x},\boldsymbol{x})&=-{1\over 288\pi^2}{\cal R}\,,
\end{split}\ee
\be
\Delta m={q^2\over 576\pi^2}{\cal R}\, .
\ee
Here ${\cal R}$ is the Ricci scalar  of the spatial metric $g_{ab}$
\be
{\cal R}={3\over 2}U^{-3}\left(U_{,a}U_{,b}-2\,UU_{,ab}
\right)\delta^{ab}\,.
\ee
                  \item{\bf  Six dimensions $D=6,n=3$}
\be\begin{split}
{\cal
G}(\boldsymbol{x},\boldsymbol{x}')&={1\over 8\pi^2}{1\over
\sqrt{UU'}}\,{1\over
|\boldsymbol{x}-\boldsymbol{x}'|^{3}}\, ,\\
{\cal
G}_\ins{div}(\boldsymbol{x},\boldsymbol{x}')&={1\over 8\pi^2}{1\over
\sqrt{UU'}}\,{1\over
|\boldsymbol{x}-\boldsymbol{x}'|^{3}}+O(|\boldsymbol{x}-\boldsymbol{x}'|)
\, ,\\
{\cal
G}_\ins{reg}(\boldsymbol{x},\boldsymbol{x})&=0\, ,
\end{split}\ee
\be
\Delta m=0\, .
\ee
\end{itemize}

Note that the obtained results are valid for the geometries which are more
general than the Majumdar-Papapetrou  spacetimes, because in the
derivation of these formulas we used the metric in the form \eq{MP} with an
arbitrary function $U$.  The Majumdar-Papapetrou spacetimes
satisfy the Einstein equations which lead to the additional constraint 
\eq{lapU} to the function $U$.


\subsection{Self-force near five dimensional Reissner-Nordstr\"{o}m black hole}

The self-force $f^{a}$ acting on a static scalar charge can be read out from the
variation of the self-energy over displacement of the charge
(see, e.g, \cite{SmithWill:1980,QuinnWald:1997,ChoTsokarosWisseman:2007} for
details)
\be
\delta E_\ins{self}=-f_{a}\delta x^a\,.
\ee
In even-dimensional asymptotically flat spacetimes of the type \eq{MP} the
self-energy vanishes and there is no corresponding self-force.
In odd-dimensional spacetimes there appears a non-trivial self-force. 

Consider a simple example of a self-force of a scalar charge near a single 
five-dimensional extremal Reissner-Nordstr\"{o}m black hole.
In this case 
\be
U=1+{M\over \rho^2} \hh {\cal R}=6{M^2\over (\rho^2+M)^3}\,.
\ee
The self-energy
\be
E_\ins{self}={q^2\over 576\pi^2}\,U^{-1}{\cal R}={q^2\over
96\pi^2}\,{M^2\rho^2\over \left(\rho^2+M\right)^4}\, .
\ee
In terms of the Schwarzschild radial coordinates \mbox{$r^2=\rho^2+M$} it reads
\be
E_\ins{self}={q^2\over
96\pi^2}\,{M^2 (r^2-M)\over r^8}\, .
\ee
Thus the only non-vanishing component of the self-force is its
radial component 
\be
f^{\rho}={q^2\over
48\pi^2}\,{M^2\rho^3(3\rho^2-M)\over \left(\rho^2+M\right)^6}\, .
\ee
The force is repulsive at far distances, vanishes at \mbox{$\rho=\sqrt{M/3}$} 
(or, equivalently, $r=2\sqrt{M}/\sqrt{3}$),
and becomes attractive at smaller radii. At the horizon $\rho=0$ it vanishes
again.


\section{Conclusions}

In the paper we discussed the problem of self-energy of a classical charged
particle in an external static gravitational field. Our main focus was on the
case of higher dimensional gravity. Classical self-energy of
pointlike charges diverges and should be properly regularized and renormalized.
Our approach is to use well established regularizations techniques of quantum
field theory to deal with this problem.
To single out divergences of the self-energy of
a pointlike charge we used the point-splitting method. This method is well
known in the quantum field theory and is convenient for our purposes. It
has been intensively used
for study of the vacuum polarization effects in black hole physics and
cosmology. We demonstrated that the application of a similar method for the
classical problem
allows one to reproduce the earlier published results of calculations
where
a special (uniformly charged shell) model of classical charged particle was
used. 

Important property of the point-splitting method is that it does not
require a special model for the charged particle, and it is easily adapted for
higher dimensional calculations. We performed calculations of the self-energy
for
a static source of a  minimally coupled scalar massless field. We showed that
the contribution of the self-energy to the proper mass of the particle has the
form
\be
\Delta m=-{1\over 2}q^2 \,{\cal G}_\ins{reg}(\BM{x},\BM{x})\, .
\ee
In other words, $\Delta m$ is identical to
\be
\Delta m=-{1\over 2}q^2\, \langle \tp^2\rangle_\ins{ren}\, ,
\ee
where $\tp$ is the properly normalized Euclidean quantum field in
$(D-1)-$dimensional space with metric $g_{ab}$.
The classical point-splitting method practically coincides with
calculations of $\langle \tp^2\rangle_\ins{ren}$ in the corresponding $(D-1)-$
quantum theory (see, e.g., \cite{Page:1982,ThompsonLemos:2009}).

To illustrate the developed method of the calculations of the self-energy, 
we applied it to a special case of a charge
in the  higher dimensional Majumdar-Papapetrou metrics, describing a set of
static extremally charged black holes in equilibrium. For such spaces the static
Green functions for a static charge are known explicitly. We showed that for
even dimensional spacetimes ($D=4,6, \ldots$) $\Delta m$ vanishes identically.
In five dimensions $\Delta m$ does not vanish and is related to the local
curvature invariants of the spatial metric. One can expect that this is a
generic property of all odd dimensional spacetimes.
A natural explanation of these
results might be the
following. On the class of spacetimes of the form \eq{MP} characterized by an
arbitrary function $U$, the corresponding operator $\hat{F}$ is
invariant under the transformation of the function $U$.
The spatial part of the Majumdar-Papapetrou metric is conformally
flat. In a flat space $\Delta m$ vanishes identically. Thus the
non-trivial value of $\Delta m$ in a `physical' space arises as a result of the
mechanism similar to the conformal anomalies. For odd dimensional
spaces with $(D-1=3,5,\ldots)$ such
anomalies vanish, and $\Delta m$ remains equal to zero. We are going to return
to this interesting problem and to discuss this mechanism in details in another
publication.

\acknowledgments

This work was partly supported  by  the Natural Sciences and Engineering
Research Council of Canada. The authors are also grateful to the
Killam Trust for its financial support.

\appendix

\section{Conformal transformation of the DeWitt
coefficients}\label{AppA}

Consider $(n+2)-$dimensional space with the conformally flat metric
\be
g_{ab}=\Omega^2 \bar{g}_{ab}\hh \sigma^c\sigma_c=2\sigma \hh
\bar{\sigma}^c\bar{\sigma}_c=2\bar{\sigma}\,.
\ee
Here  ${\sigma}^a\equiv g^{ab}{\sigma}_{,b}$ and $\bar{\sigma}^a\equiv
\bar{g}^{ab}\bar{\sigma}_{,b}$. The metric $\bar{g}_{ab}$ is a flat metric.
Therefore
\be
\bar{\sigma}_{a;b}=\bar{g}_{ab}\,.
\ee

We can express $\sigma$ in terms of $\bar{\sigma}$ and its derivatives. The
result is
\be\begin{split}
\sigma=\bar{\sigma}\Big[
\Omega^2-\Omega\Omega_{;a}\bar{\sigma}^a
+{1\over 12}(4\Omega\Omega_{;ab} \hskip 0.6cm &\\
+4\Omega_{;a}
\Omega_{;b}
-\Omega_{;c}\Omega^{;c}\bar{g}_{ab})\bar{\sigma}^a\bar{\sigma}^b
\Big]&+\dots\\
=\bar{\sigma}\,\Omega(\boldsymbol{x})\Omega(\boldsymbol{x}')
\Big[1 +{1\over
12\,\Omega^{2}}(-2\Omega\Omega_{;ab}
 &\\
+4\Omega_{;a}\Omega_{;b}
-\Omega_{;c}\Omega^{;c}\bar{g}_{ab})\bar{\sigma}^a\bar{\sigma}^b\Big]&+\dots
\end{split}\ee
For the determinant  $\Delta^{1/2}(\boldsymbol{x},\boldsymbol{x}')$ we have
\be\begin{split}
\Delta^{1/2}&=1+{1\over
12}{\cal R}_{ab}\sigma^a\sigma^b+\dots\\
&=1+{1\over
12\Omega^2}\Big[
-n\Omega\Omega_{;ab}+2n\Omega_{;a}\Omega_{;b}
\\
&-\left(\Omega\Omega_ {;c
}^{;c}+(n-1)\,\Omega^{;c}\Omega_{;c}\right)\bar{g}_{ab}
\Big]\bar{\sigma}^a\bar{\sigma
}^b+\dots
\end{split}\ee
Here we took into account
\be\begin{split}
{\cal R}_{ab}=\bar{{\cal R}}_{ab}+{1\over\Omega^2}\Big[
-n\Omega\Omega_{;ab}+2n\Omega_{;a}\Omega_{;b} & \\
-\left(\Omega\Omega_ {;c
}^{;c}+(n-1)\,\Omega^{;c}\Omega_{;c}\right)\bar{g}_{ab}
\Big] \,.&
\end{split}\ee
\be
{\cal R}={1\over\Omega^2}\bar{{\cal R}}-{n+1\over\Omega^4}\left[
2\Omega\Omega_{;c}^{;c}+(n-2)\Omega^{;c}\Omega_{;c} \right]\,.
\ee
and that in our case  $\bar{g}_{ab}$ is flat and, hence,
$\bar{R}_{ab}=0$.
Here, on the right hand side of these equations all the covariant derivatives
$\Omega_{;ab}, \Omega_{;a}$ etc. are defined in accordance with the flat metric
$\bar{g}_{ab}$.

The first DeWitt coefficients corresponding to the operator
\be
\hat{F}=\nabla^a\nabla_a+V
\ee
are
\be
a_0(\boldsymbol{x},\boldsymbol{x}')=1\,.
\ee
\be\begin{split}
a_1(\boldsymbol{x},\boldsymbol{x}')&=V+{1\over 6}{\cal R}+\dots
\end{split}\ee

\section{DeWitt coefficients in Majumdar-Papapetrou
spacetimes}\label{AppB}

The  expressions \eq{Gdiv4}--\eq{Gdiv6} for the UV divergent terms of
the scalar
Green function have been derived for a generic curved spacetimes. Now
we apply
these results to the class of Majumdar-Papapetrou metrics.

In the case of the  metric \eq{MP} we have
\be
\Omega=U^{1/n}\,,
\ee
\be\begin{split}
{\cal R}&={1\over 4\xi_{n+2}}U^{-2-{2\over n}}\left(
U_{,a}U_{,b}-2\,UU_{,ab}\right)\delta^{ab}\,,\\
\xi_{n+2}&={n\over
4(n+1)}\,.
\end{split}\ee
\be\begin{split}
V&=-\xi_{n+2}\,{\cal R}
\\
&=-{1\over 4}U^{-2-{2\over n}}\left(
U_{,a}U_{,b}-2\,UU_{,ab}\right)\delta^{ab}\,.
\end{split}\ee
\be\begin{split}
&a_1(\boldsymbol{x},\boldsymbol{x})={1\over 6}{\cal R}+V=\Big({1\over
6}-\xi_{n+2} \Big){\cal R}\,.
\end{split}\ee

\begin{itemize}
\item

In five dimensions $n=2$, $\Omega=U^{1/2}$, $\xi_4=1/6$

\be
a_1(\boldsymbol{x},\boldsymbol{x})\equiv 0\,,
\ee
\be
{\Delta^{1/2}\over
\sigma}={1\over\Omega\bar{\sigma}\Omega'}+{1\over 36}{\cal R}+\dots\,,
\ee
and
\be
{\cal R}={3\over
2}U^{-3}\left(U_{,a}U_{,b}-2\,UU_{,ab}
\right)\delta^{ab}\,.
\ee

\item

In six dimensions  $n=3$, $\Omega=U^{1/3}$, $\xi_5=3/16$
\be\begin{split}
a_1&=-{1\over 48}{\cal R}\,,
\end{split}\ee
\be
{\cal R}=-{4\over\Omega^4}\left[
2\Omega\Omega_{;c}^{;c}+\Omega^{;c}\Omega_{;c} \right]\,,
\ee
\be\begin{split}
\Delta^{1/2}
=1+{1\over 12\Omega^2}\Big[
-3\Omega\Omega_{;ab}+6\Omega_{;a}\Omega_{;b}
&\\
-(\Omega\Omega_ {;c}^{;c}
+2\,\Omega^{;c}\Omega_{;c})\bar{g}_{ab}
\Big]\bar{\sigma}^a\bar{\sigma}^b &+\dots\,,
\end{split}\ee
\be\begin{split}
\Delta^{1/2}\left[
{1\over
(2\sigma)^{3/2}}\,{a}_0+{1\over
2(2\sigma)^{1/2}}\,{a}_1\right]&\\=
{1\over
\Omega^{3/2}(2\bar{\sigma})^{3/2}{\Omega'}^{3/2}}
&+O({\bar{\sigma}}^{1/2})
\\=
{1\over U^{1/2}(2\bar{\sigma})^{3/2}{U'}^{1/2}}
&+\dots\,.
\end{split}\ee

In all odd-dimensional $(n=2,4,\dots)$ Majumdar-Papapetrou
spacetimes the DeWitt coefficients
\be
a_{n/2}(\boldsymbol{x},\boldsymbol{x}')=0\,.
\ee
These coefficients appear as the factors before
{$\ln|\boldsymbol{x}-\boldsymbol{x}'|$} in the Hadamard representation.
This is why there are no logarithmic divergences in the static Green
functions in these spacetimes.


\end{itemize}

\end{document}